# ERSFQ 8-bit Parallel Arithmetic Logic Unit

A. F. Kirichenko, I. V. Vernik, M. Y. Kamkar, J. Walter, M. Miller, L. R. Albu, and O. A. Mukhanov

***Abstract*— We have designed and tested a parallel 8-bit ERSFQ arithmetic logic unit (ALU). The ALU design employs wave-pipelined instruction execution and features modular bit-slice architecture that is easily extendable to any number of bits and adaptable to current recycling. A carry signal synchronized with an asynchronous instruction propagation provides the wave-pipeline operation of the ALU. The ALU instruction set consists of 14 arithmetical and logical instructions. It has been designed and simulated for operation up to a 10 GHz clock rate at the 10-kA/cm$^2$ fabrication process. The ALU is embedded into a shift-register-based high-frequency testbed with on-chip clock generator to allow for comprehensive high frequency testing for all possible operands. The 8-bit ERSFQ ALU, comprising 6840 Josephson junctions, has been fabricated with MIT Lincoln Lab's 10-kA/cm$^2$ SFQ5ee fabrication process featuring eight Nb wiring layers and a high-kinetic inductance layer needed for ERSFQ technology. We evaluated the bias margins for all instructions and various operands at both low and high frequency clock. At low frequency, clock and all instruction propagation through ALU were observed with bias margins of +/-11% and +/-9%, respectively. Also at low speed, the ALU exhibited correct functionality for all arithmetical and logical instructions with +/-6% bias margins. We tested the 8-bit ALU for all instructions up to 2.8 GHz clock frequency.**

***Index Terms*—energy-efficient computation, superconducting, digital processing, SFQ, ERSFQ.**

## I. Introduction

SUPERCONDUCTING digital technology is progressing from venerable Rapid Single Flux Quantum (RSFQ) [1], [2] to energy-efficient SFQ [3]-[7] and adiabatic logics [8], [9]. These RSFQ successors are now considered to be a basis for the next-generation low-power circuit technology needed for future high energy-efficient data centers, supercomputers [10]-[12], and embedded classical control modules for quantum computers [13]. One of the most promising and practical energy efficient SFQ logic is ERSFQ, which retains all advantages of RSFQ including the well-developed circuit libraries [3], [14], [15]. ERSFQ logic is one of two integrated circuit technologies chosen for the implementation of superconducting processors in C3 project [16].

The Arithmetic Logic Unit (ALU) is a key processing module of the central processing unit (CPU) of a computer. Since the ALU is the most actively switching CPU module, it largely defines the performance and drives the architectural choices of the entire processing unit.

The research is based upon work supported in part by the Office of the Director of National Intelligence (ODNI), Intelligence Advanced Research Projects Activity (IARPA), via contract W911NF-14-C0090.

A. F. Kirichenko, I. V. Vernik, M. Y. Kamkar, J. Walter, M. Miller and O. A. Mukhanov are with HYPRES, Elmsford, NY 10523 USA (e-mail: alex@hypres.com)

L. R. Albu was with IBM T.J. Watson Research Center, Yorktown Heights, NY 10598 USA (email: remus.albu@gmail.com)

To date, the reported superconductor ALU designs were implemented using RSFQ logic following bit-serial, bit-slice, and parallel architectures.

The bit-serial designs have the lowest complexity; however, their latencies increase linearly with the operand lengths, hardly making them competitive for implementation in 32-/64-bit processors [17], [18]. Bit-serial ALUs were used in 8-bit RSFQ microprocessors [19]-[24], in which an 8 times faster internal clock is still feasible. As an example, an 80 GHz bit-serial ALU was reported in [25].

In order to alleviate the high-clock requirements and long latencies of bit-serial design while keeping moderate hardware complexity, a bit-slice ALU architecture was implemented [26], [27]. Typically in such designs, 4-bit parallel slices were run serially, reducing the internal clock requirements, but the delay induced by carry signal rotation limits the overall ALU clock frequency.

Parallel ALUs are capable of delivering a competitive performance for 32-/64-bit processors. Parallel architecture implemented in RSFQ enables a very high throughput, however the latency still can be large due to deep execution pipelines. In order to reduce latency, an asynchronous wave-pipeline timing was proposed. In such a timing scheme, an operation can start immediately when two independent operands arrive. No clock pulse is needed to move to the next stage. A clock pulse follows the data to reset cells for the next data wave. Using such a design approach, a 20 GHz 8-bit parallel ALU was demonstrated based on a Kogge-Stone parallel adder amended with a rich set of instructions [28], [29]. However, the circuit hardware complexity was significant with large circuit area, making it impractical for integration within a microprocessor. In attempt to simplify the hardware, a parallel-prefix sparse-tree ALU was reported [30]. To date, the implemented parallel ALUs are not practical for the construction of a compact high-performance microprocessor.

In this paper, we report the first implementation of a parallel ERSFQ 8-bit ALU. In order to reduce complexity and enable highly compact integrated circuit design, a hardware-efficient ripple carry architecture was employed. This design was combined with the asynchronous wave-pipelined timing architecture. This allowed us to preserve high throughput and prevent the latency increase that is typical in straightforward ripple carry designs.

## II. Design

### A. Architecture

The goal of our project was to demonstrate a fully functional ERSFQ parallel 8-bit CPU placed on a single 5x5-mm chip fabricated in MIT Lincoln Lab SFQ5ee process [31]. Therefore, the chosen architecture needed to be very compact. The most compact parallel ALU architecture is based on a simple ripple-carry adder. RSFQ is a sequential logic family, so in order to achieve high throughput (i.e., clock speed) we have employed the wave pipeline architecture [32], [33].

The ERSFQ ripple-carry wave-pipelined adder has been described and demonstrated in [14]. It comprises two rows of half adders with asynchronous carry propagation. The first row of half adders performs XOR and AND operations on the operands, then the second row produces SUM and CARRY.

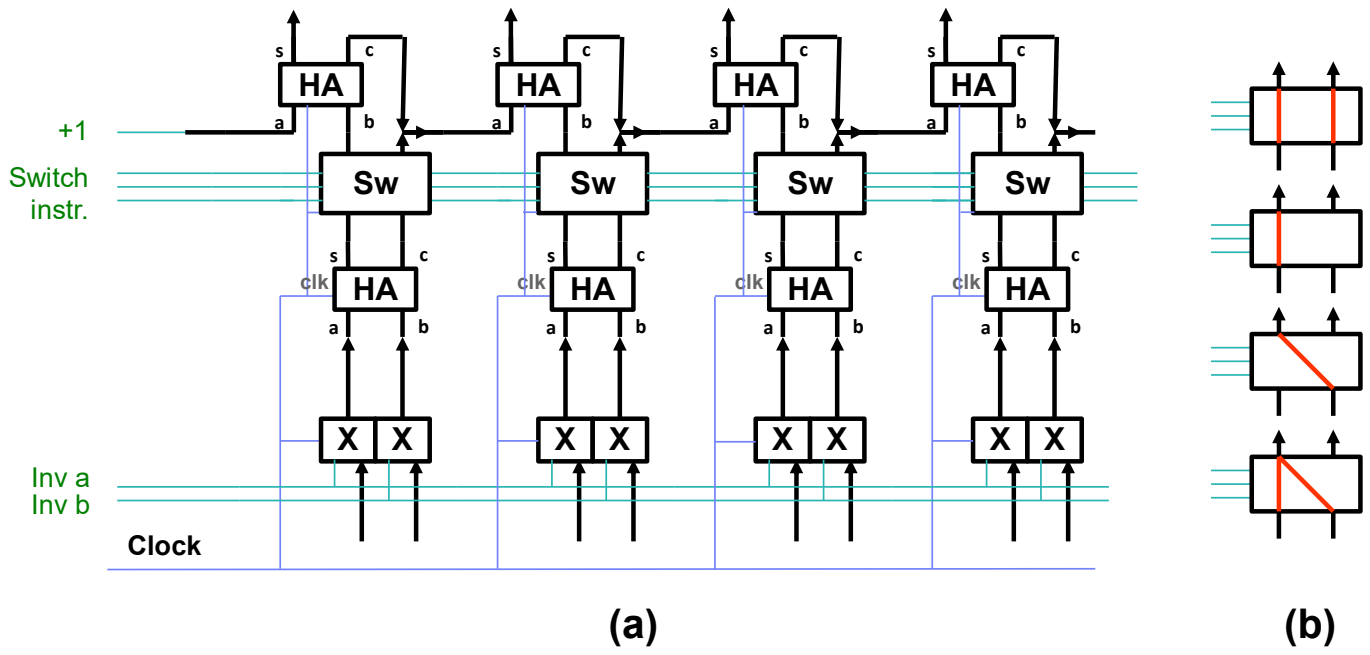

Fig. 1. A fragment of the ALU block diagram (a) and the instruction switch configurations (b). Here, "X" stands for an XOR cell, "HA" – a half adder, and "Sw" is an instruction switch.

In order to make a simple ALU out of such an adder, we separated these half adder rows with reconfigurable 2x2 switches (Fig. 1a). Depending on the switch configuration (Fig. 1b), the result of this simple ALU can be OR, AND, XOR, or ADD. The wave pipeline ensures synchronization of the CARRY propagation along the upper half adders with the 3-bit instruction code propagation along the instruction switches (Fig. 1). We have slowed down horizontal clock signal propagation so the clock is following data, forming a wave. As a result, as soon as the first bit slice has calculated the carry the next instruction can be sent for execution, producing an asynchronous (wave) pipeline.

In order to expand the instruction set (Table I), we have added XOR gates at the input to serve as controllable inverters. Together with unoccupied input of the LSB second row half adder ("+1" in Fig. 1a), the inversion of the input operands allows such operations as subtraction and comparison, in addition to inversion itself and many logical combinations.

Three slices of the ALU (1842 Josephson junctions) were successfully simulated at physical level and optimized with newly available superconductor circuit simulator PSCAN2 [34].

### B. *Instruction switch*

The instruction decoding switch is the circuit that turns a ripple-carry parallel adder into an ALU (Fig. 1a). Its functionality is to commutate SUM and CARRY outputs from the first row of half adders to the second row of half adders. There are five useful configuration states of the switch: ZERO, AND, XOR, OR, and ADD (Fig. 1b). Thus, one instruction takes three control lines. With two argument inversions and one "*carry in*" ("*+1*") control line, there are in total 6 instruction control lines in the ALU (Table I).

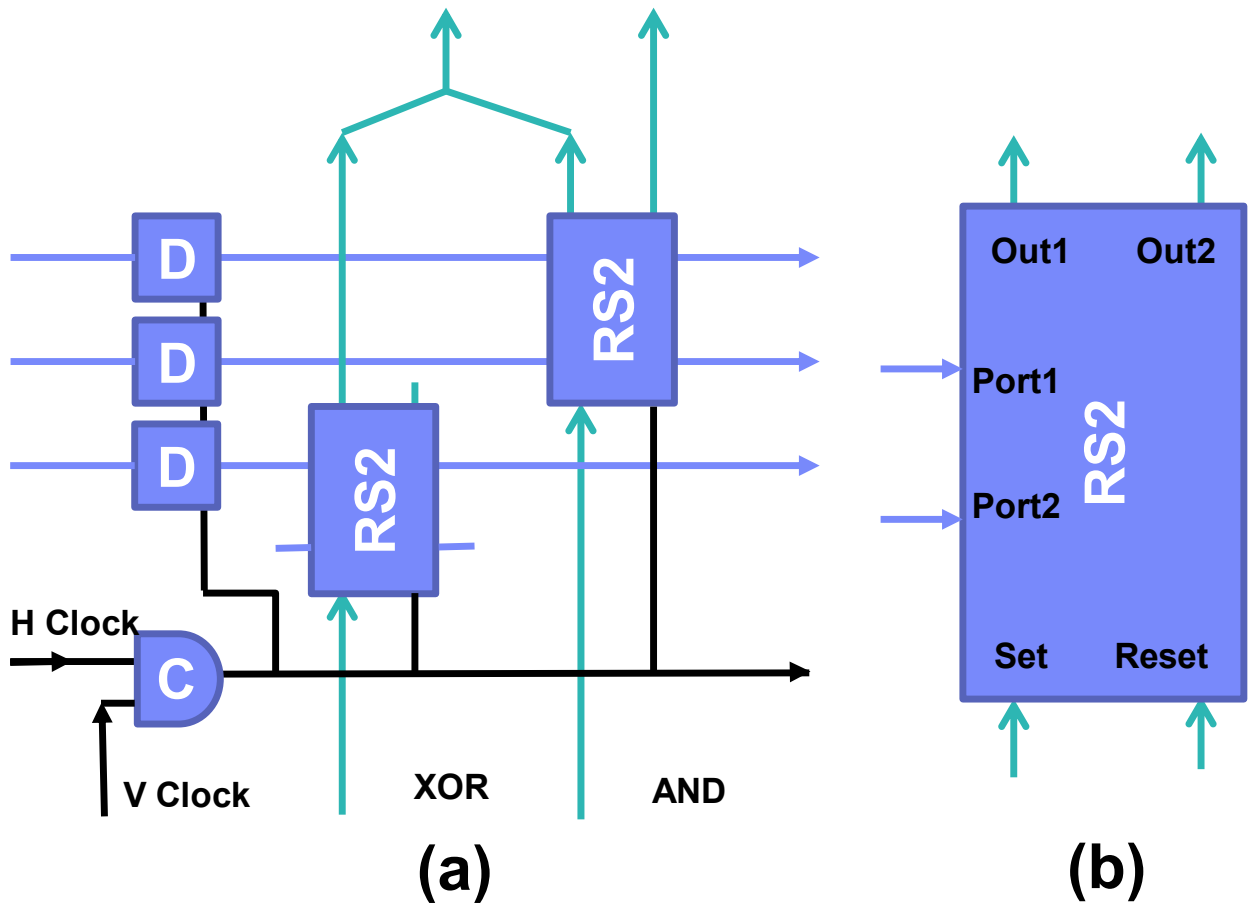

Fig. 2. Block-diagram of the instruction switch (a) and the dual-port RS flip flop symbolic notation (b).

TABLE I
ALU INSTRUCTION SET

| | +1 | XOR | AND | CARRY | Inv. A | Inv. B |
|---|---|---|---|---|---|---|
| Zero | 0 | 0 | 0 | 0 | 0 | 0 |
| XOR | 0 | 1 | 0 | 0 | 0 | 0 |
| NXOR | 0 | 1 | 0 | 0 | 1 | 0 |
| AND | 0 | 0 | 1 | 0 | 0 | 0 |
| NAND | 0 | 1 | 1 | 0 | 1 | 1 |
| OR | 0 | 1 | 1 | 0 | 0 | 0 |
| NOR | 0 | 0 | 1 | 0 | 1 | 1 |
| A+B | 0 | 1 | 0 | 1 | 0 | 0 |
| A-B | 1 | 1 | 0 | 1 | 0 | 1 |
| B-A | 1 | 1 | 0 | 1 | 1 | 0 |
| NOT (A) | 0 | 1 | 0 | 0 | 1 | 0 |
| NOT (B) | 0 | 1 | 0 | 0 | 0 | 1 |
| A=B | 1 | 1 | 0 | 0 | 0 | 1 |
| Inc A | 1 | 1 | 0 | 0 | 0 | 0 |

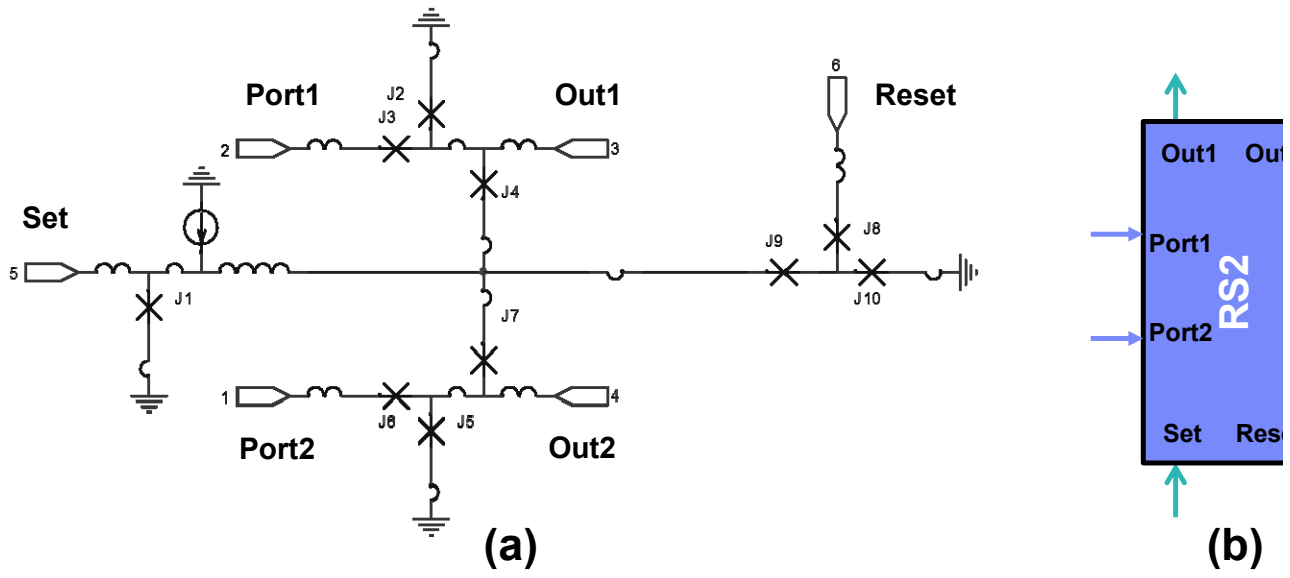

Fig. 3. Schematics of the dual-port RS flip-flop cell (a) and its symbolic notation (b).

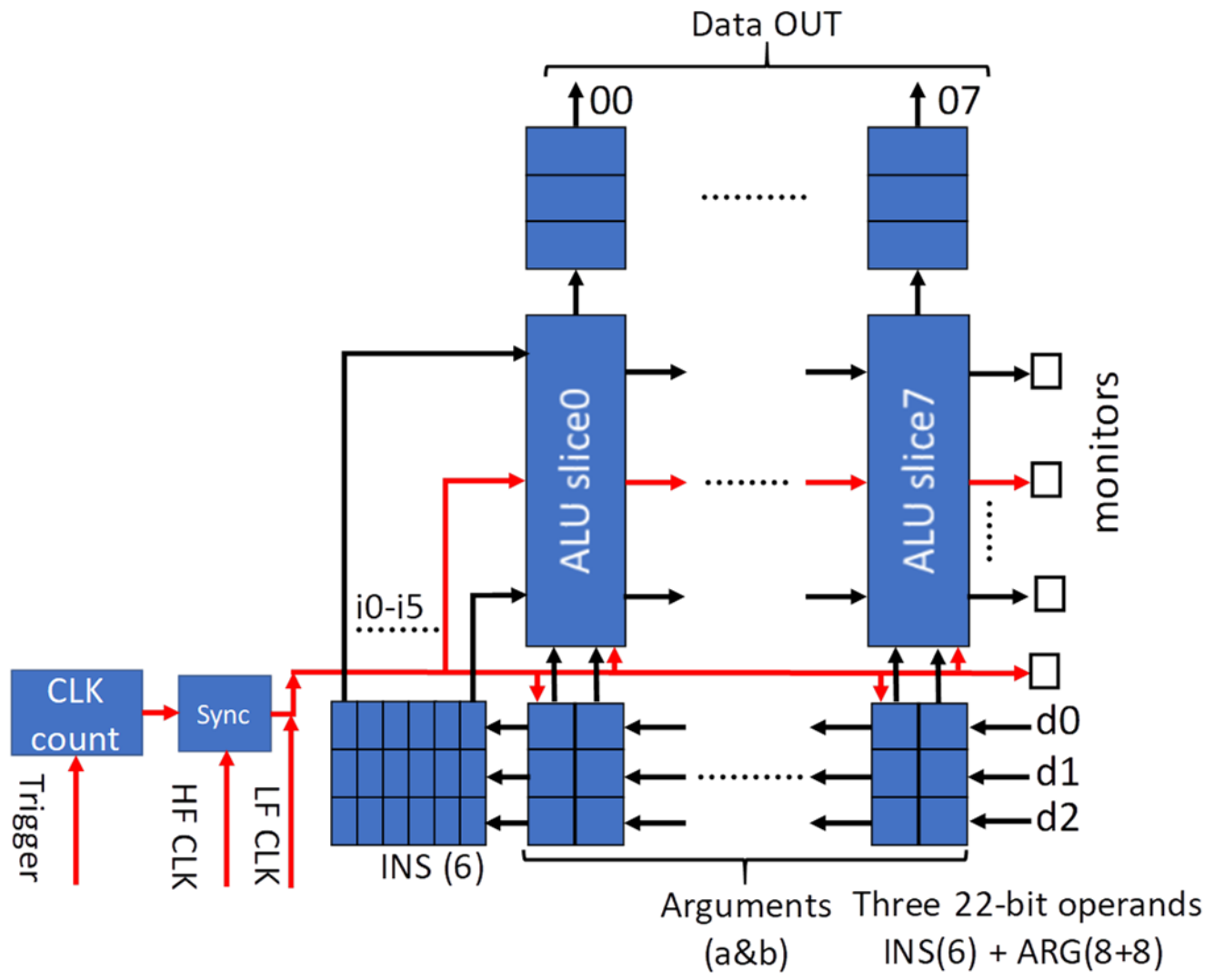

Fig. 4. Block-diagram of the testbed for high-speed testing of the 8-bit ALU.

The instruction switch (Fig. 2a) comprises two dual-port RS flip-flops [35], [36] (Fig. 2b); one is for commutating the XOR input signal and the other is for commutating the AND input signal. The schematics of the RS2 cell is shown in Fig. 3. Every clock cycle a clock signal clears both RS2 cells, arriving at their respective "Reset" ports. Depending on which port the destructive readout signal arrives, the input signal emits from either the Out1 or Out2 output ports.

The instruction is propagated between ALU slices over passive transmission lines (PTLs) in sync with a wave-pipeline signal by means of D flip-flop latches. The wave pipeline synchronization is provided by Muller's C-element, which synchronizes horizontal and vertical clock signal propagation.

### C. *ALU Integrated with High Frequency Test-Bed*

The built-in high-frequency test bed is a natural feature of sequential logic. Because of its internal memory, RSFQ has natural capacity for scan-chains and buffers. Therefore testing a complex parallel RSFQ circuit at high speed is a relatively easy task when employing an on-chip testbed [37], [38].

Fig. 4 shows a block diagram of the high-frequency test bed used for testing the ALU. The 8-bit ALU operand consists of two 8-bit arguments (*a0-a7* and *b0-b7*) and a 6-bit instruction (*i0-i5*), making the input register 22-bits long. Here we used a three-word-deep buffer, meaning we can test at high speed up to three operands sequentially. The input register buffer consists of a 3x22 matrix of dual-port D flip-flops (D2) [35] connected in series horizontally via one port and vertically via the second port. This comprises a scan chain that can serially load (at low speed) three 22-bit input operands and upload them in parallel (at high speed) into the circuit under test (the ALU).

The high-speed clock is generated by a clock generator. After the input operands are loaded, at the trigger signal the HF clock counter [36] generates exactly four SFQ pulses. These pulses are being synchronized with an external HF generator, allowing the varying of the clock frequency. During these four clocks cycles, three consecutive operands get executed and the result gets stored at the output buffer. The results can thus be read out from the output buffer at low speed and analyzed. This approach allows for the exhaustive testing of ALU exceeding the room temperature interface limit speed and also without exploitation of very expensive high-frequency equipment.

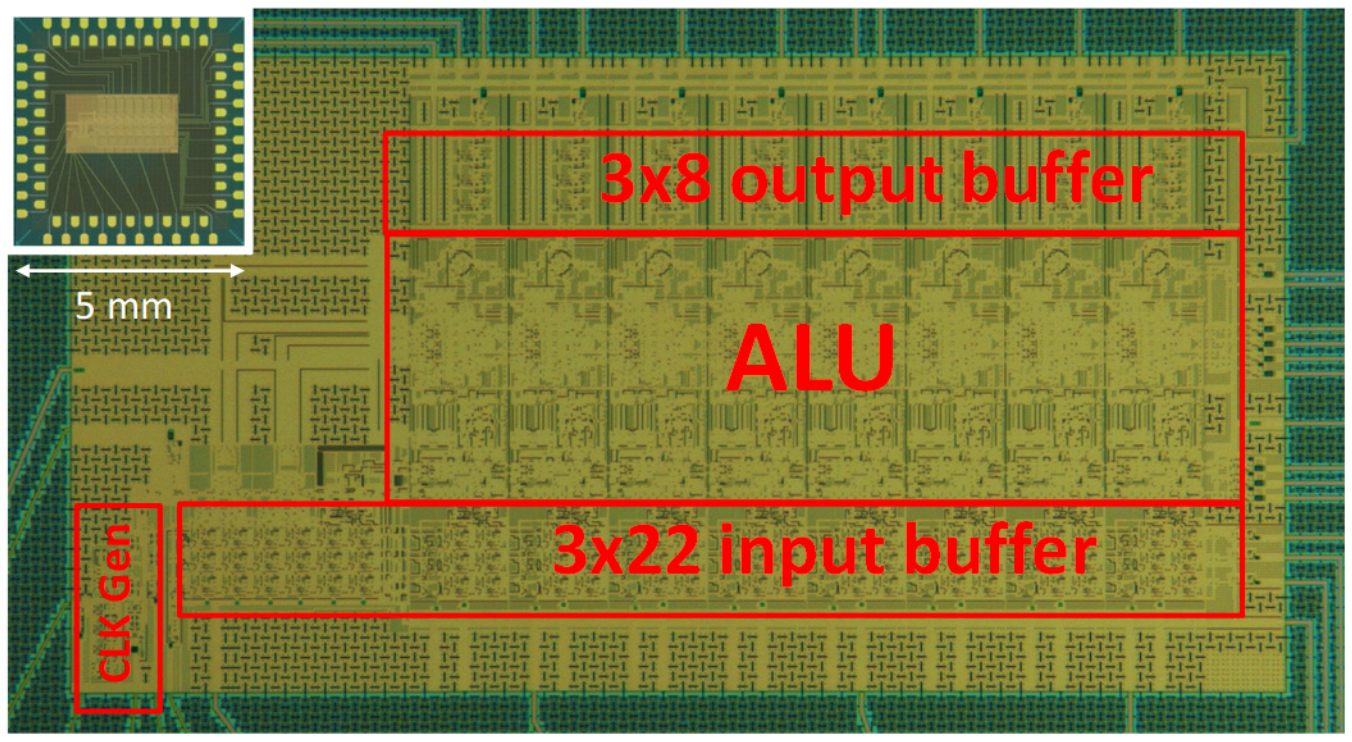

Fig. 5. Micrograph of the ALU with high frequency testbed. The inset shows the whole 5 x 5 mm$^2$ chip (zoom out).

### D. *Layout*

The ALU integrated into the high-speed testbed was designed for and fabricated in MIT-LL SFQ5ee 10-kA/cm$^2$ fabrication process [31] featuring eight Nb wiring layers and a high kinetic inductance layer (HKIL). The HKIL allows the placement of large ERSFQ bias inductors under the circuit area, substantially reducing physical size of ERSFQ circuits.

Fig. 5 shows a micrograph of the 8-bit ERSFQ ALU placed in the high-speed testbed on a 5 x 5-mm$^2$ chip, with ALU and components of the test bed marked. The whole circuit comprises 6840 Josephson junctions, including ERSFQ bias JJs (1008) and the testbed (920). The 8-bit ALU occupies an area of 1.6 x 0.5 mm$^2$. The ALU is designed with slice pitch of 200 μm, matching the pitch of the designed Register File of the CPU. The whole ALU has a single common bias current line.

We have simulated and optimized three ALU bit-slices using new PSCAN2 simulator [34]. We have also performed LVS verification/extraction with Cadence, followed by the circuit simulation (back annotation) with Cadence Spectre for verification of the ALU functionality with the extracted circuit parameters. This close-to-physical-layout simulation is taking into account the whole periphery, such as the extracted ERSFQ biasing network and the chip wiring.

## III. Experimental Results

### A. *Prescreening Tests*

The experimental evaluation of a chip of such complexity takes a few steps including prescreening. Because the yield of our fabrication process has not reached the industrial standard yet, we had to come up with some prescreening procedure in order to quickly discard a defective chip from a poorly yielded wafer. The most critical and prone-to-defect object in the 9-layer process [31] is a via connecting all metal layers. In our (ERSFQ) case, this is a via between the HKIL (on the bottom of the stack) and one of the wiring layers connecting to a bias Josephson junction (so-called bias inductor). By simply measuring the critical current of the ERSFQ bias line, we could detect disconnected bias inductors and discard such chips. Even a single defective bias inductor out of thousands would lead to functionality failure, although sometimes the circuit would still work with smaller than expected operational margins. So if the critical current of the ALU bias line was significantly less than the designed value, that chip would not function properly and was discarded.

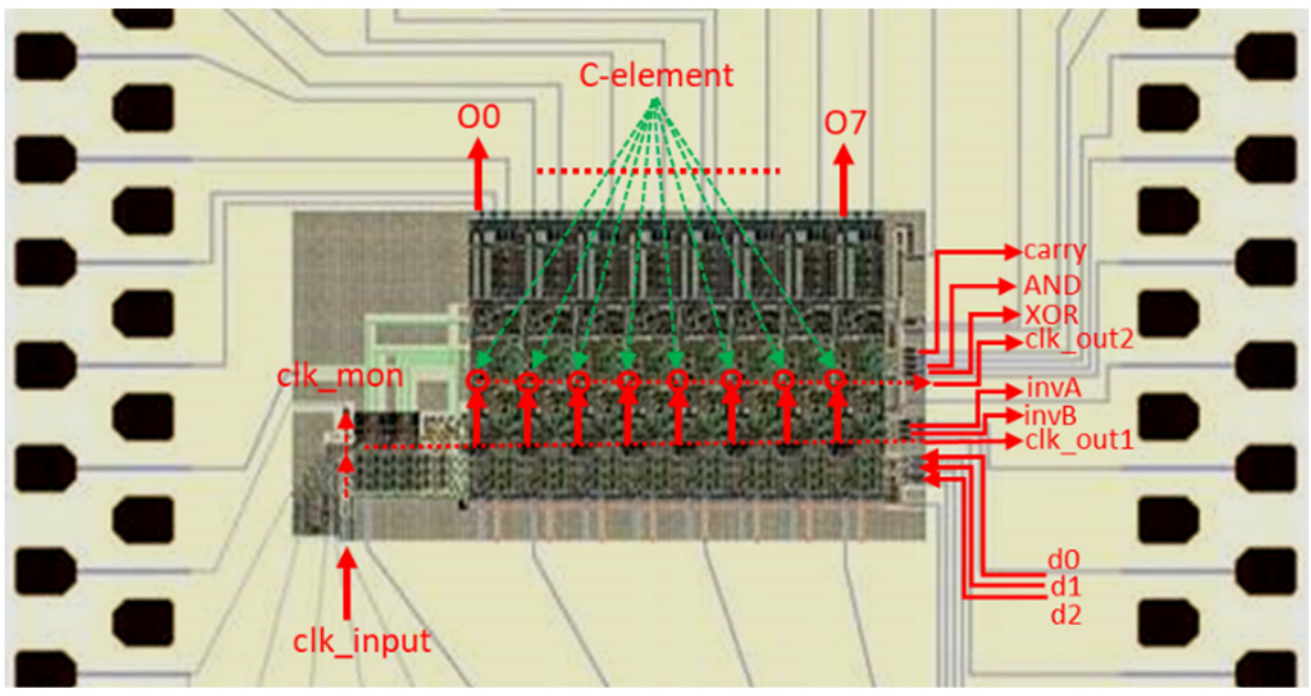

Fig. 6. The ALU in HF testbed on a 5-mm chip with marked inputs/outputs.

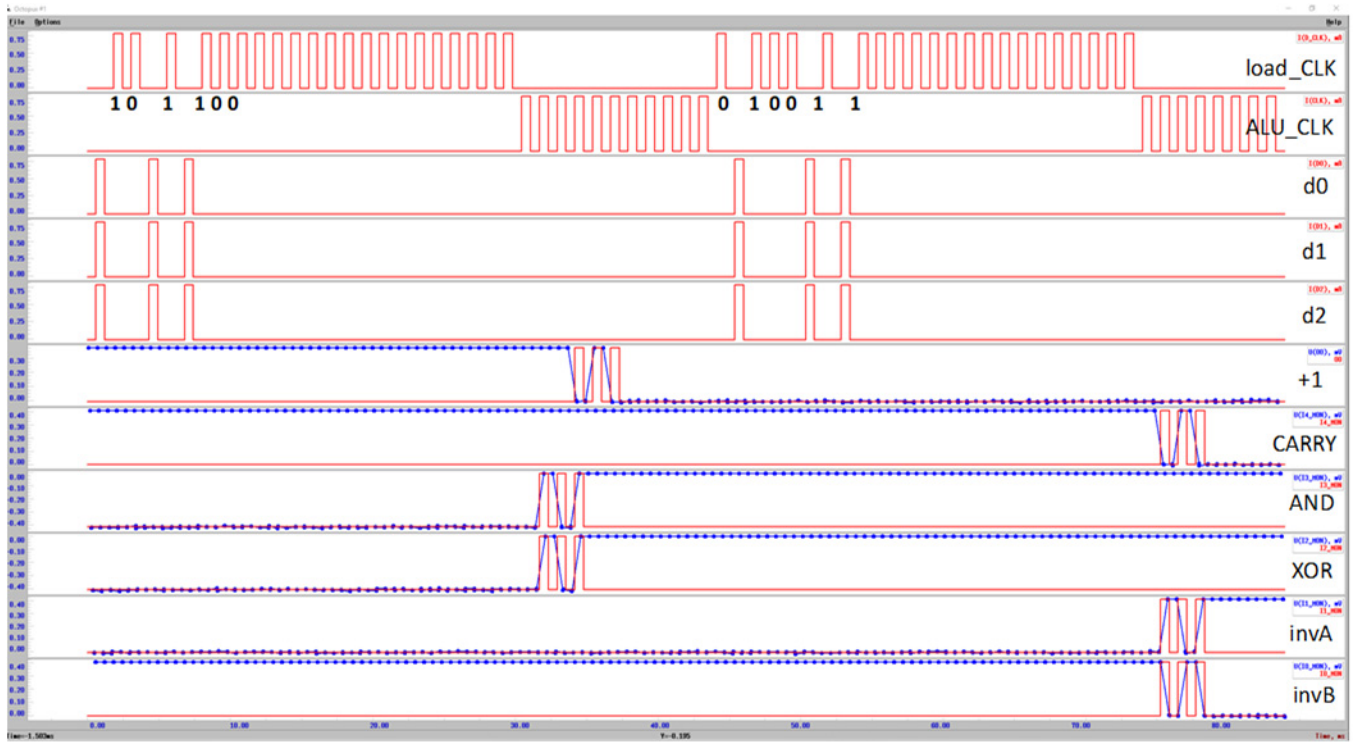

Fig. 7. Correct propagation of all instructions through ALU with bias margins of ± 9.6%. The red traces indicate input signals applied and expected outputs. The blue traces are the instruction monitor's outputs being observed with the toggle-type SFQ/dc converters (Fig. 6).

After analog prescreening is done, we start the digital testing. On the first run we observe clock propagation along the ALU by three clock monitors specifically designed for diagnostics purpose. The clock monitors were placed at the entrance to the ALU (*clk_mon* in Fig. 6), at the exit from the last ALU slice (*clk_out1*), and at the exit from the wave-pipeline synchronization chain (*clk_out2*). The last clock monitor shows whether the whole wave-pipeline synchronization works properly. The presence of this signal indicates that the clock has successfully passed through the chain of eight Muller C-elements and passive transmission lines (PTLs).

After the chip passes the clock chain prescreening procedure, we start the actual digital testing by verifying instruction propagation through the ALU (Fig. 7). This test indicates not only that all ALU instruction switches are operational, but also that the testbed is functioning properly. An instruction is loaded as the first 6 bits of a 22-bit operand. Three consecutive operands are loaded via ports d0-d2 in Fig. 6. Five ("*invB", "invA", "XOR", "AND",* and *"CARRY"*) out of six instruction control lines can be directly monitored at the ALU outputs marked in Fig. 6. The instruction control line ("*+1*") can be monitored via LSB output ("*O0*").

Fig. 7 shows the correct behavior of all 6 instruction control lines. This test can be performed at low speed only, as none of the monitors are buffered except "*+1*". The monitor of "*+1*" (*carry in*) instruction control line is an output of the first slice (LSB) and is buffered, so it is shifted by two clock cycles relatively to the other five monitors.

## *B. Functionality and High-Speed Test*

Once the chip has successfully passed the prescreening process, we start exhaustive evaluation of the ALU at low and high speed and measure its bias operating margins.

Fig. 8 shows a block diagram of the high-speed on-chip test approach [37] we used. Here we used the OCTOPUX test system [39] for biasing the ALU and the on-chip testbed, providing control and input signals, and monitoring the outputs. The low-speed and the high-speed tests look almost identical, with the exception of the high-frequency clock trigger signal that is applied in the high-speed test mode.

In order to run the operation execution, first the input buffer has to be loaded with three 22-bit operands (Fig. 4). Then we apply low-speed ALU clock signal to get three consecutive 8-bit results at the parallel output.

In high-speed mode, before applying the clock, a "HF trigger" signal (Fig. 8) is sent to the on-chip clock counter [37], producing four SFQ pulses at an extremely high speed defined by the delay in a single merger. These pulses go to the synchronization block comprising a set of latches clocked by the external HF generator. Thus, for every initial signal, the on-chip clock generator produces four clock pulses at the externally given frequency.

One clock pulse is required for the ALU to run a wave-pipelined operation and three pulses are for writing the results to a 3x8-bit output buffer. The next three clock pulses read the 8-bit results out.

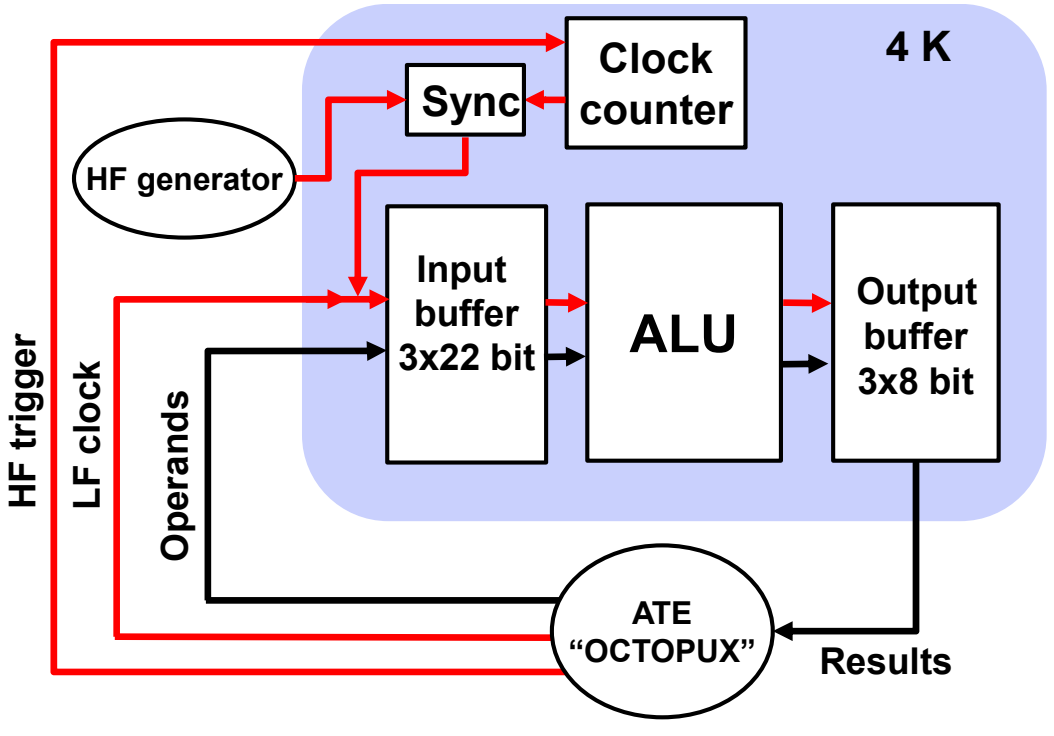

Fig. 8. Block diagram of high-speed test of the ALU.

Although the on-chip buffer memory depth limits the number of operands that can be tested in each cycle, this high speed on-chip approach allows unambiguous verification of every operand execution between any two other operands.

## C. Test Results

We successfully tested the ERSFQ 8-bit ALU embedded in the high frequency testbed for the instruction set listed in Table I with various arguments at low speed. The operational bias current margins were within ± 6.5 % (from $I_{low}$ = 413 mA to $I_{high}$ = 470 mA). Fig. 9 shows the correct ALU operation for *ADD, AND,* and *XOR*. Fig. 9a shows two results of the instruction *A*+*B*: at *A* = 29, *B* = 141 and at *A* = 13, *B* = 72. Fig. 9b shows results for instructions *AND* and *XOR* on arguments *A* = 63 and *B* = 240. These are output patterns only.

In order to illustrate the functionality of the testbed, we present in Fig. 10 full test patterns with the same operands as in Fig. 9. The HF clock signal (Fig. 8) was set to 2.8 GHz. The top 4 traces are the operand loading input signals (d0-d2 and the serial clock). The next (fifth) trace is the ALU clock, which performs execution of the operands and reads out the results. The sixth trace, designated as CLK_TR, is the "HF trigger" signal. This is where the low- and the high-speed patterns differ. The final 8 traces comprise the parallel outputs of the output buffer (actually measured and digitized).

The dashed line in Fig. 10 divides the experiment into the low- and the high-speed mode with the same input operands. The high-speed test differs from the low-speed test by presence of "HF trigger" pulse (marked CLK_TR in Fig. 10). Naturally, with the same inputs for low- and high-speed tests, both halves in Fig. 10 show the same output (*O0-O7*). The only other difference between the two halves of the picture is the phase shift by 4 clocks. At low-speed clock, as we discussed above, the first result should appear on the fifth ALU clock signal. While in high-speed mode, the first output occurs on the first (5-4=1) low-speed clock pulse. This is clearly seen in Fig. 10.

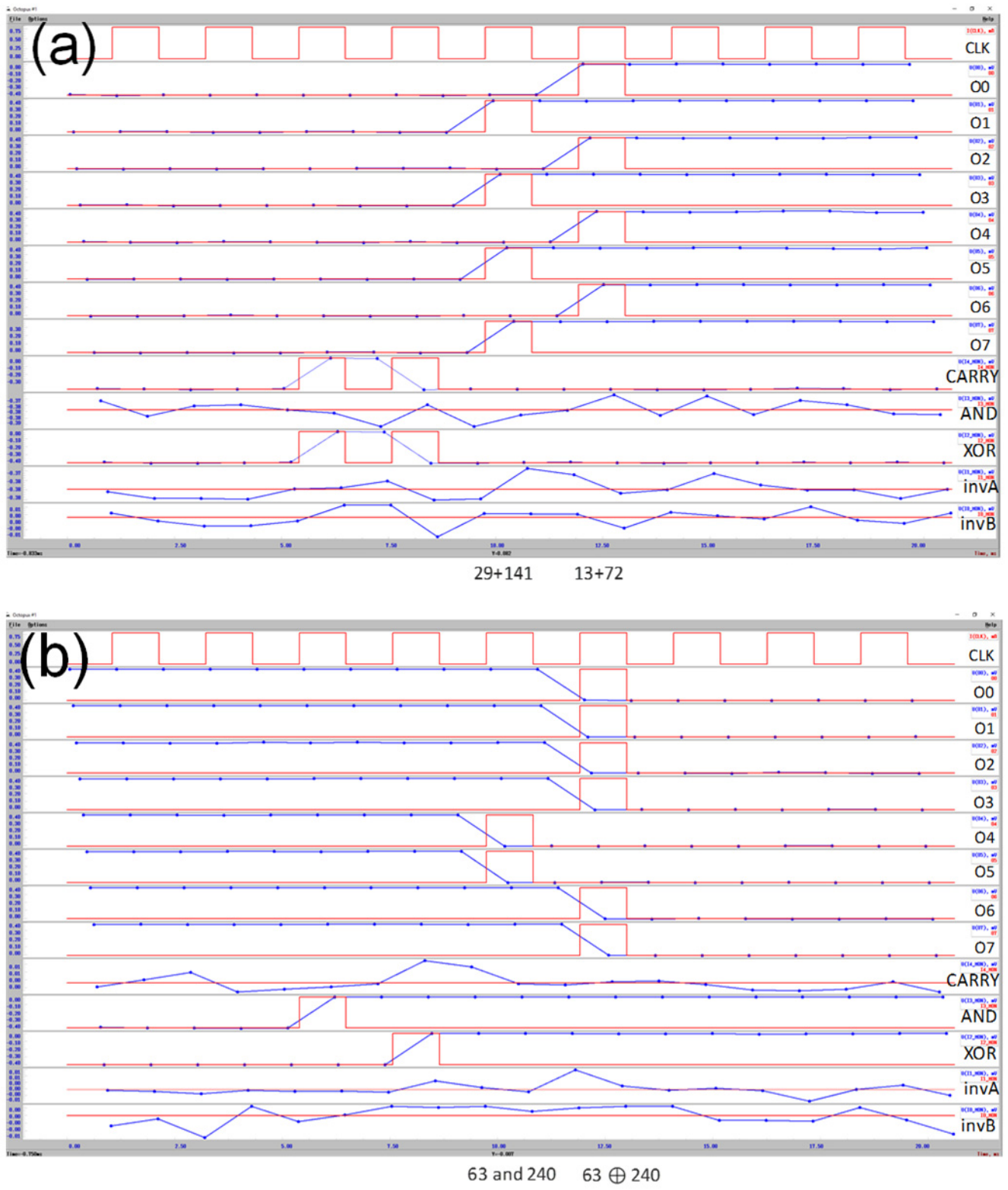

Fig. 9. Correct low-speed operation of the ALU with bias margins of ± 6.5% for operands (29+141) and (13+72) (a) and for (63 AND 240) and (63 XOR 240) (b). The pattern is with the output traces only. The top trace is a clock signal. The rest are the outputs O0-O7 and the instruction monitors.

TABLE II
EXPERIMENTALLY MEASURED ALU FUNCTIONAL BIAS MARGINS (LF)

| Test type | $I_{low}$ (mA) | $I_{high}$ (mA) | Full margins (+/- %) | Margins under critical current [a] (+/- %) |
|---|---|---|---|---|
| Clock propagation | 398 | 502 | 11.6 | 3.9 |
| Instruction propagation | 405 | 491 | 9.6 | 3.0 |
| Operations execution | 413 | 470 | 6.5 | 2.0 |

[a] The upper margin is the critical current of the whole bias network

## IV. Discussion

Table II lists low frequency bias margins for clock and all instruction propagation through the ALU, as well as margins for the ALU's correct functionality for all arithmetical and logical instructions. The last column in Table II shows the margins for the most energy-efficient operation [40] of the ALU, i.e. in zero static power dissipation mode.

The reduced measured bias margins (both full and most energy-efficient) when compared to simulated ones can be understood by possible fabrication issues affecting bias current delivery via the HKIL. It is worth noting that if a particular HKIL bias inductor was defective causing bias delivery to the ERSFQ cell to fail, the ERSFQ circuit would still work with the required bias current redistributed from the adjacent cells under overbiased conditions. Strong experimental indications of such a behavior were observed, with chip-to-chip variation of $I_C$ of the ERSFQ bias network, and the shifting of bias margins above ERSFQ $I_C$ even for simple clock propagation.

The same, we believe, goes for the explanation why did the ALU with the testbed operate at 2.8 GHz only, while it was designed to work at up to 10-GHz clock frequency.

## V. Conclusion

We have designed and tested a wave-pipeline energy-efficient parallel ALU based on ERSFQ logic.

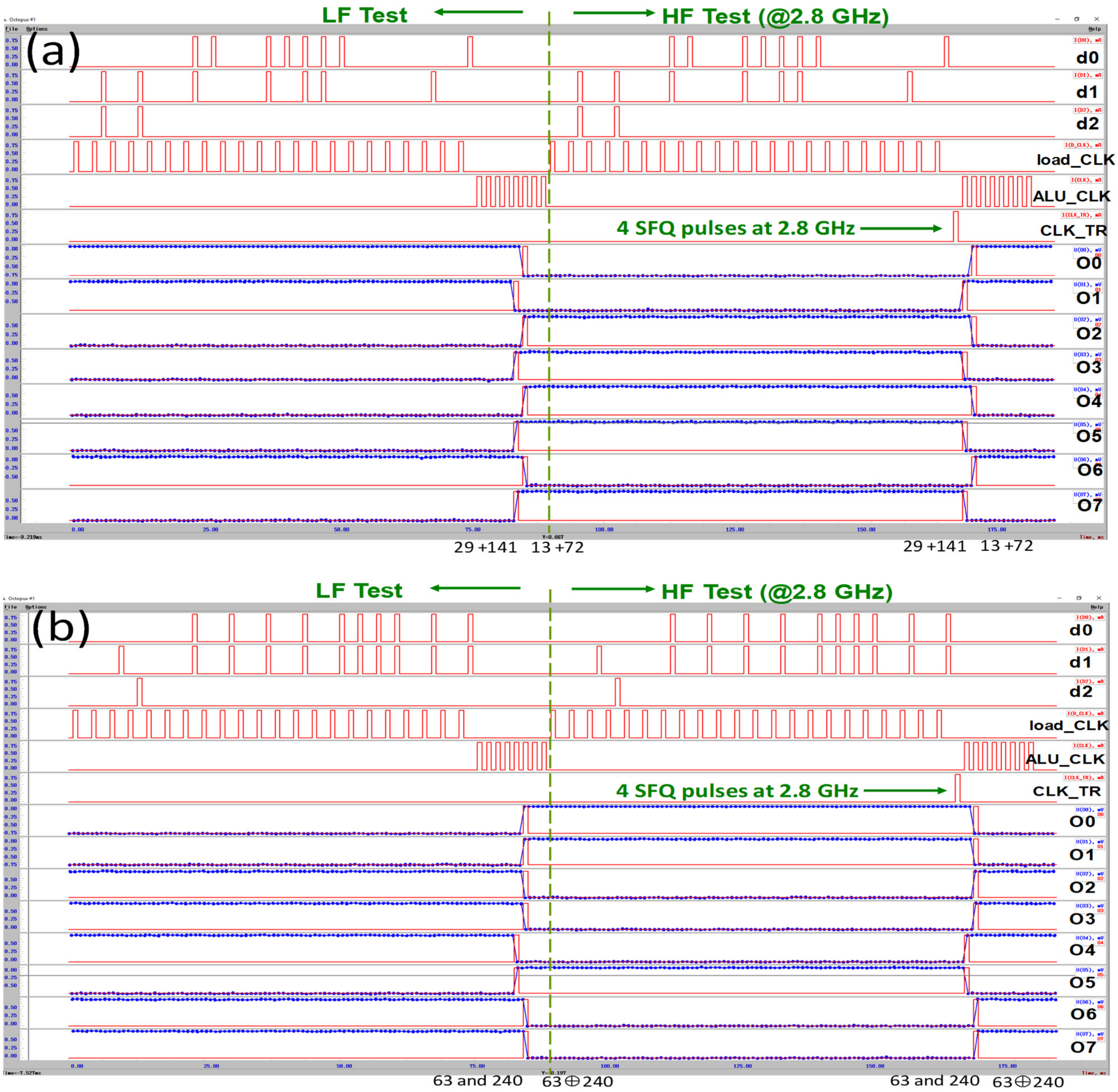

Fig. 10. The full pattern with low-speed (left half) and high-speed 2.8 GHz (right half) test for operands (29 + 141), (13 + 72) (a) and (63 AND 240), (63 XOR 240) (b). The top 6 traces are the input signals loading operands and applying clock. The bottom 8 traces (O0-O7) are the output.

The compact ripple carry design enables us to fit the entire 8 bit ALU with 14 instructions into a relatively small 1.6 x 0.5 $mm^2$ area with a 200-µm slice pitch. Such a compact design is made possible by using ERSFQ functional cells with internal memory instead of synthesizing the design using elementary gates and latches. For area-efficient signal distribution, we used a combination of passive transmission lines between ALU slices and JTLs for routing signals within individual slices. This modular approach is also convenient for implementing so-called "current recycling" scheme in ALU biasing.

In order to preserve high throughput, we used wave pipelined timing that allows the next operation to start immediately when two operands arrive. This timing scheme also allows us to reduce ALU latency, since each data wave propagates asynchronously throughout the entire ripple carry circuit without the need of clock. The clock wave is only used to reset ERSFQ cell before the next data wave.

The ALU circuit comprises 6840 Josephson junctions and was fabricated using a 10-kA/$cm^2$ MIT-LL SFQ5ee fabrication process with eight Nb wiring layers and HKIL. The 8-bit ALU was subsequently integrated into an 8-bit ERSFQ microprocessor [41], which is placed on a 5 x 5 $mm^2$ chip.

Our ERSFQ 8-bit ALU operated successfully for all arithmetical and logical instructions up to 2.8 GHz clock frequency. To facilitate unambiguous digital testing, the ALU circuit was embedded into a high-frequency testbed with on-chip SFQ clock generator. This approach allowed us to verify any possible operand combination at both low and high speed. We measured bias margins for all instructions and various operand combinations at both low and high frequency clock.

## Acknowledgment

The authors thank the MIT Lincoln Lab foundry team for chip fabrication, M. Denneau and S. V. Rylov for valuable discussions, and S. Holmes and M. Manheimer for support.

The views and conclusions contained herein are those of the authors and should not be interpreted as necessarily representing the official policies or endorsements, either expressed or implied, of the ODNI, IARPA, or the U.S. Government.

## References

[1] K. K. Likharev and V. K. Semenov, "RSFQ logic/memory family: A new Josephson-junction technology for sub-terahertz-clock-frequency digital systems," IEEE Trans. Appl. Supercond., vol. 1, no. 1, pp. 3–28, Mar. 1991.

[2] O. A. Mukhanov, V. K. Semenov, and K. K. Likharev, "Ultimate performance of the RSFQ logic circuits," *IEEE Trans. Magn.*, vol. MAG-23, no. 2, pp. 759-762, Mar. 1987.

[3] D. Kirichenko, S. Sarwana, A. Kirichenko, "Zero static power dissipation biasing of RSFQ circuits," *IEEE Trans. Appl. Supercon*., vol. 21, pp.776-779, Jun. 2011.

[4] O. A. Mukhanov, "Energy-efficient Single Flux Quantum technology" *IEEE Trans. Appl. Supercond.,* vol. 21, pp.760-769, Jun. 2011.

[5] M. H. Volkmann, S. Sahu. C. Fourie, O. A. Mukhanov, "Implementation of energy efficient single flux quantum digital circuits with sub-aJ/bit operation," *Supercond. Sci. Technol*. vol. 26, Jan. 2013, Art. No. 015002.

[6] M. Tanaka, M. Ito, A. Kitayama, T. Kouketsu, A. Fujimaki, "18-GHz, 4.0-aJ/bit operation of ultra-low-energy Rapid Single-Flux-Quantum shift registers," *Jap. J. Appl. Phys.* vol.51*,* 2012, Art. No. 053102.

[7] Q. Herr, A. Herr, O. Oberg, A. Ioannidis, "Ultra-low-power superconductor logic," *J. Appl. Phys.,* vol. 109, 2011. Art. No. 103903.

[8] N. Takeuchi, D. Ozawa, Y. Yamanashi, N. Yoshikawa, "An adiabatic quantum flux parametron as an ultra-low-power logic device," *Supercond. Sci. Technol.* vol. 26, 2013, Art. No. 035010.

[9] N. Takeuchi, Y. Yamanashi, N. Yoshikawa, "Adiabatic quantum-flux-parametron cell library adopting minimalist design," *J. Appl. Phys*., vol. 117, no. 17, 2015, Art. no. 173912.

[10] D. S. Holmes, A. L. Ripple, M. A. Manheimer, "Energy-efficient superconducting computing – power budgets and requirements," *IEEE Trans. Appl. Supercond.,* vol. 23, Jun. 2013, Art. no. 1701610.

[11] S. Nishijima, S. Eckroad, A. Marian, K. Choi *et al.*, "Superconductivity and the environment: a Roadmap," *Supercond. Sci. Technol.* vol. 26, 2013, Art. no. 113001.

[12] D. S. Holmes, A. M. Kadin, M. W. Johnson, "Superconducting computing in large-scale hybrid systems," *Computer*, vol. 48, pp. 34-42, Dec. 2015.

[13] R. McDermott, M. G. Vavilov, B. L. T. Plourde, F. K. Wilhelm, P. J. Liebermann, O. A. Mukhanov, T. A. Ohki, "Quantum-classical interface based on single flux quantum digital logic," *Quantum Sci. Technol.*, vol.3, no. 2, Jan. 2018, Art. no. 024004.

[14] A. F. Kirichenko, I. V. Vernik, J. A. Vivalda, R. T. Hunt and D. T. Yohannes, "ERSFQ 8-bit parallel adders as a process benchmark," *IEEE Trans. Appl. Supercond.*, vol. 25, Jun. 2015, Art. no. 1300505.

[15] A. F. Kirichenko, I. V. Vernik, O. A. Mukhanov and T. A. Ohki, "ERSFQ 4-to-16 Decoder for Energy-Efficient RAM," *IEEE Trans. Appl. Supercond.*, vol. 25, Jun. 2015, Art. no. 1301304.

[16] M. A. Manheimer, "Cryogenic computing complexity program: Phase 1 introduction", *IEEE Trans. Appl. Supercond*., vol.25, Jun. 2015, Art. no. 1301704.

[17] K. Takahashi, S. Nagasawa, H. Hasegawa, K. Miyahara, H. Takai, and Y. Enomoto, "Design of a superconducting ALU with a 3-input XOR gate," *IEEE Trans. Appl. Supercond*., vol. 13, no. 2, pp. 551–554, Jun. 2003

[18] J. Y. Kim, S. Kim, and J. Kang, "Construction of an RSFQ 4-Bit ALU with half adder cells," *IEEE Trans. Appl. Supercond*., vol. 15, no. 2, pp. 308– 311, Jun. 2005

[19] M. Dorojevets, P. Bunyk, and D. Zinoviev, "FLUX chip: Design of a 20-GHz 16-bit ultrapipelined RSFQ processor prototype based on 1.75-μm LTS technology," IEEE Trans. Appl. Supercond., vol. 11, no. 1, pp. 326–332, Mar. 2001.

[20] M. Tanaka et al., "A single-flux-quantum logic prototype microprocessor," in Proc. IEEE ISSCC Dig. Tech. Papers, Feb. 2004, vol. 1, pp. 298–529.

[21] Y. Yamanashi et al., "Design and implementation of a pipelined bit-serial SFQ microprocessor, CORE1β," *IEEE Trans. Appl. Supercond*., vol. 17, no. 2, pp. 474–477, Jun. 2007.

[22] A. Fujimaki et al., "Bit-serial single flux quantum microprocessor CORE," *IEICE Trans. Electron*., vol. E91-C, pp. 342–349, Mar. 2008.

[23] M. Tanaka et al., "Design and implementation of a pipelined 8 bit-serial single-flux-quantum microprocessor with cache memories," *Supercond. Sci. Technol*., vol. 20, no. 11, pp. S305–S309, Nov. 2007.

[24] Y. Nobumori et al., "Design and implementation of a fully asynchronous SFQ microprocessor: SCRAM2," *IEEE Trans. Appl. Supercond*., vol. 17, no. 2, pp. 478–481, Jun. 2007.

[25] Y. Ando, R. Sato, M. Tanaka, K. Takagi and N. Takagi, "80-GHz Operation of an 8-Bit RSFQ Arithmetic Logic Unit," in *2015 15th International Superconductive Electronics Conference (ISEC)*, Nagoya, 2015, pp. 1-3.

[26] G. Tang, K. Takata, M. Tanaka, A. Fujimaki, K. Takagi and N. Takagi, "4-bit Bit-Slice Arithmetic Logic Unit for 32-bit RSFQ Microprocessors," *IEEE Trans. Appl. Supercond.*, vol. 26, no. 1, Jan. 2016, Art no. 1300106.

[27] G. Tang, P. Qu, X. Ye and D. Fan, "Logic Design of a 16-bit Bit-Slice Arithmetic Logic Unit for 32-/64-bit RSFQ Microprocessors," *IEEE Trans. Appl. Supercond.*, vol. 28, no. 4, Jun. 2018, Art no. 1300305.

[28] T. Filippov et al., "8-bit asynchronous wave-pipelined RSFQ arithmetic logic unit," *IEEE Trans. Appl. Supercond*., vol. 21, no. 3, pp. 847–851, Jun. 2011.

[29] T. Filippov et al., "20 GHz operation of an asynchronous wave-pipelined RSFQ arithmetic-logic unit," *Phys. Procedia*, vol. 36, pp. 59–65, 2012.

[30] M. Dorojevets, C. Ayala, N. Yoshikawa, and A. Fujimaki, "8-bit asynchronous sparse-tree superconductor RSFQ arithmetic-logic unit with a rich set of operations," *IEEE Trans. Appl. Supercond*., vol. 23, no. 3, Jun. 2013, Art. ID 1700104.

[31] S. K. Tolpygo, V. Bolkhovsky, T. J. Weir, A. Wynn, D. E. Oates, L. M. Johnson, and M. A. Gouker, "Advanced Fabrication Processes for Superconducting Very Large Scale Integrated Circuits," *IEEE Trans. Appl. Supercond.*, vol. 26, Jun. 2016, Art. no. 1100110.

[32] L. Cotten, "Maximum rate pipelined systems," in Proc. AFIPS Spring Joint Comput. Conf., 1969.

[33] M. Dorojevets, C. Ayala and A. Kasperek, Data-Flow Microarchitecture for Wide Datapath RSFQ Processors: Design Study, *IEEE Trans. Appl. Supercond.*, vol. 21, issue 3, Jun. 2011, pp. 787–791, 12022856

[34] PSCAN2. Available at http://www.pscan2sim.org/index.html

[35] S. V. Polonsky, V. K. Semenov, A. F. Kirichenko, "Single flux, quantum B flip-flop and its possible applications," *IEEE Trans. Appl. Supercond.*, vol. 4, pp. 9-16, March 1994.

[36] K. Fujiwara, H. Hoshina, J. Koshiyama, and N. Yoshikawa, "Design and component test of RSFQ packet decoders for shift register memories," *Physica C*, vol. 378–371, pp. 1475–1480, 2002.

[37] Z. J. Deng, N. Yoshikawa, S. R. Whiteley, and T. Van Duzer, "Data-Driven Self-Timed RSFQ High Speed Test System," *IEEE Trans. Appl. Supercond.*, vol. 7, pp. 3830–3833, Dec. 1997.

[38] A.F. Kirichenko, O. A. Mukhanov, and A.I. Ryzhikh, "Advanced On-Chip Test Technology for RSFQ Circuits," *IEEE Trans. on Applied Supercond.*, vol. 7, No. 2, pp. 3438-3441, Jun. 1997.

[39] D. Y. Zinoviev and Y. A. Polyakov, "Octopux: an advanced automated setup for testing superconductor circuits," *IEEE Trans. Appl. Supercond.*, vol. 7, no. 2, pp. 3240-3243, Jun. 1997.

[40] C. Shawawreh, D. Amparo, J. Ren, M. Miller, M. Y. Kamkar, A. Sahu, A. Inamdar, A. F. Kirichenko, O. A. Mukhanov, and I. V. Vernik, "Effects of adaptive dc biasing on operational margins in ERSFQ circuits," *IEEE Trans. Appl. Supercond.,* vol. 27, Jun. 2017, Art. no. 1301606.

[41] A. Kirichenko et al., "ERSFQ 8-bit CPU design," ASC 2016 conference, Denver, CO, USA, Sept. 4–9, 2016, report 4EOr2B-02.